\documentclass[review]{elsarticle}
\usepackage{lineno,hyperref}
\modulolinenumbers[5]
\journal{Journal of \LaTeX\ Templates}

\begin{document}
\begin{frontmatter}
\title{Mott transition in two-band fermion model with on-site Coulomb repulsion}
\tnotetext[mytitlenote]{Mott transition in two-band fermion model with on-site Coulomb repulsion}
\author{Igor N.Karnaukhov}
\address{G.V. Kurdyumov Institute for Metal Physics, 36 Vernadsky Boulevard, 03142 Kiev, Ukraine}
\fntext[myfootnote]{karnaui@yahoo.com}



\begin{abstract}
We provide analytical and numerical solution of the two band fermion model with on-site Coulomb at half filling.
In limiting cases for generate bands and one flat band, the model reduces to the Hubbard and Falicov-Kimball models, respectively. We have shown that the insulator state emerges at half filling due to hybridization of fermions of different bands with momenta k and k$+\pi$. Such hybridization breaks the conservation of the number of particles in each band, the Mott transition is a consequence of spontaneous symmetry breaking. A gap in the spectrum is calculated depending on the magnitude of on-site Coulomb repulsion and the width of the band for the chain, as well as for square and cubic lattices. The proposed approach allows us to describe the formation of the gap in the fermion spectra in the Hubbard and Falicov-Kimball models within the framework of the same mechanism for an arbitrary dimension of the system.
\end{abstract}

\begin{keyword}
\texttt Hubbard model  \sep Falicov-Kimball model \sep Mott transition
%\pacs{71.10.Fd}
%\pacs{71.27.+a}
\end{keyword}
\end{frontmatter}
%\linenumbers

\section{Introduction}
The Hubbard model is the most popular model in condensed matter physics for investigation of the metal-insulator Mott transition, the Falicov-Kimball model also describes mixed valence phenomena. The metal-insulator phase transition is realized in the models at half filling occupation, while physical properties of insulator state are different, so the insulator phase in the Falicov-Kimball model is an excitonic insulator (see \cite{DG1,DG2} for details). The (1+1)D Hubbard model has been solved exactly by Lieb and Wu \cite{LW}, since models with non-degenerate bands are in principle exactly unsolvable, the Falicov-Kimball chain is not integrable.

Approximate methods such as the approximation of the variational cluster \cite{1}, the approximation of the dynamic cluster \cite{1a}, as well as numerical methods such as the theory of the dynamic mean field \cite {2,3}, the theory of the cellular dynamic mean field \cite {3a} are used to calculate the phase diagram of the Hubbard model, the magnitude of the gap in the spectrum of fermions.
Unfortunately, there is a wide range of $ {U}_c $ values ($  {U}_c $  is the minimal value of the on-site interaction at which the gap in the spectrum opens) calculated by various methods and approximations: according to numerical calculations $  {U}_c = 2 $  \cite {1} and $  {U }_c = 6 $ \cite {3a} for a square lattice; in a cubic lattice, analytical calculations lead to $  {U}_c = 0.87$ and $ 2.1$ for the Hubbard-III and Hubbard-I approximations.
And finally, we are not aware of the scenario of the formation of a gap in the Hubbard model (as in the case of the formation of a gap in superconductors). There is currently no consensus and understanding the Mott transition. This is primarily due to the lack of controlled approximations in the case of strong interaction in low-dimensional lattices.

The article reports on a study of the formation of gap in the Hubbard and Falicov-Kimball models in these seemingly different models. We believe that the nature of the Mott gap does not depend on the dimension of the system, namely, such, as in the (1+1)D Hubbard model, it was formed without breaking the translational symmetry in the nonmagnetic state.
The renormalization of the band spectrum with allowance for the interaction between electrons calculated perturbatively by itself cannot cause a metal-insulator transition, the phase transition is result of breaking spontaneous symmetry (see for example \cite{IK1,IK2}). In the paper, we use this idea to consider the two-band model with on-site Coulomb repulsion at half filling occupation.  The formation of a gap in the spectrum of fermions in chain and square, cubic lattices is considered, and the same scenario of the Mott transition in the Hubbard and Falicov-Kimball models is demonstrated. These models can be considered as limiting cases of the considered two-band  model.

\section{Half-filled Hubbard model for various dimensions}

The Hamiltonian of the Hubbard model has the well-known form ${\cal H}_{Hub} ={\cal H}_{0} +{\cal H}_{int} $
\begin{eqnarray}
&&{\cal H}_{0} = - \sum_{<i,j>} \sum_{\sigma =\uparrow,\downarrow} a^\dagger_{i,\sigma} a_{j,\sigma}-\mu \sum_{j} \sum_{\sigma =\uparrow,\downarrow}n_{j,\sigma},\nonumber\\&&
{\cal H}_{int}={U} \sum_{j}\left(n_{j,\uparrow}-\frac{1}{2}\right)\left(n_{j,\downarrow}-\frac{1}{2}\right),
\label{eq-Hub}
\end{eqnarray}
where $a^\dagger_{j,\sigma}$ and $a_{j,\sigma}$ are the Fermi operators that determine the electrons on a site \emph{j} ($\sigma=\uparrow,\downarrow$ is the spin of the electron),  $n_{j,\sigma} =a^\dagger_{j,\sigma}a_{j,\sigma}$ denote the density operators, $\mu$ is the chemical potential, ${U}$ is magnitude of on-site Coulomb repulsion, summation in ${\cal H}_{0}$ runs over adjacent lattice sites.

\begin{figure}[tp]
     \centering{\leavevmode}
\begin{minipage}[h]{.315\linewidth}
\center{
\includegraphics[width=\linewidth]{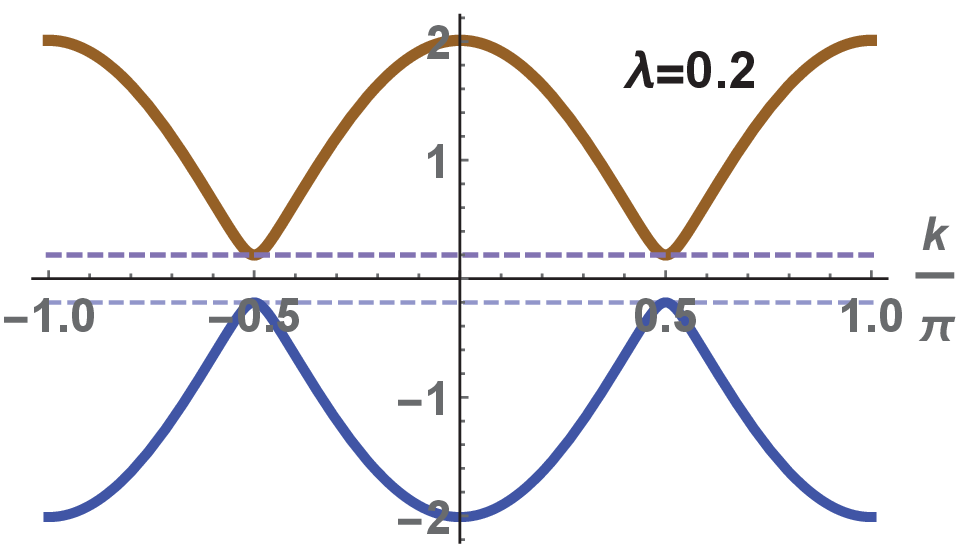} a)\\
                  }
\end{minipage}
\centering{\leavevmode}
\begin{minipage}[h]{.315\linewidth}
\center{
\includegraphics[width=\linewidth]{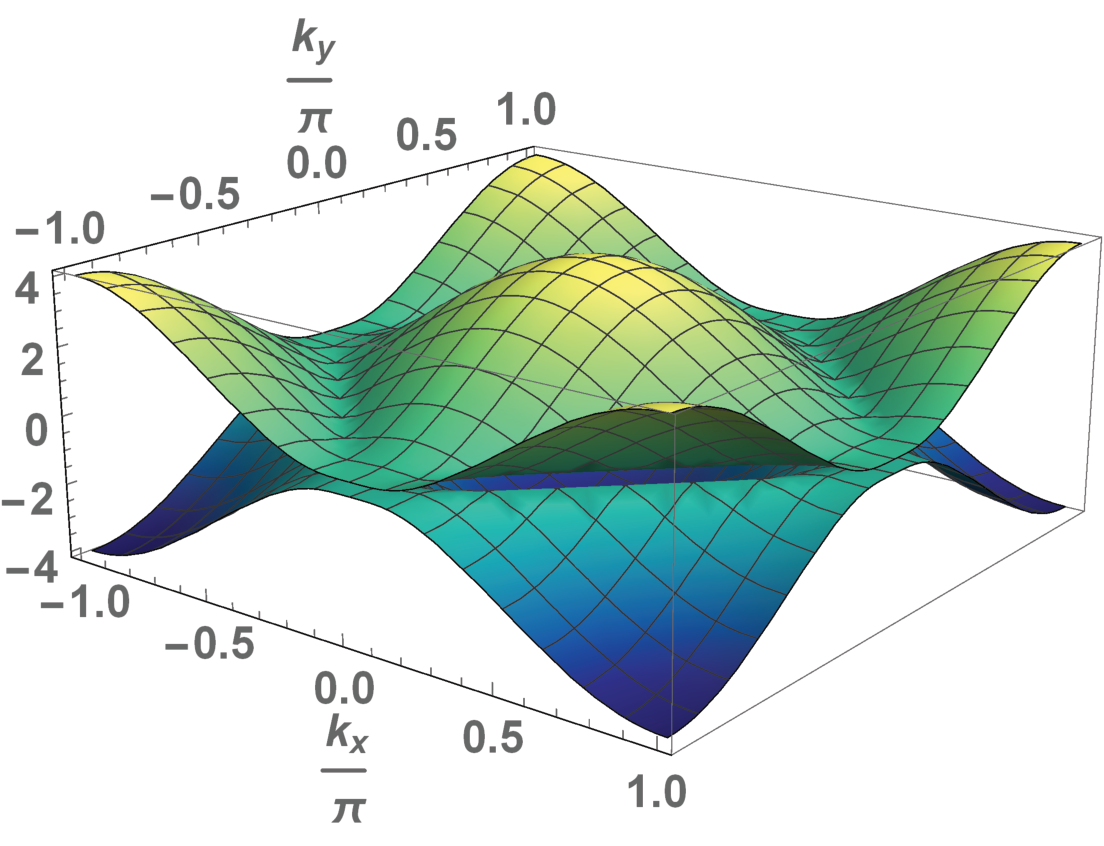} b)\\
}
\end{minipage}
\centering{\leavevmode}
\begin{minipage}[h]{.315\linewidth}
\center{
\includegraphics[width=\linewidth]{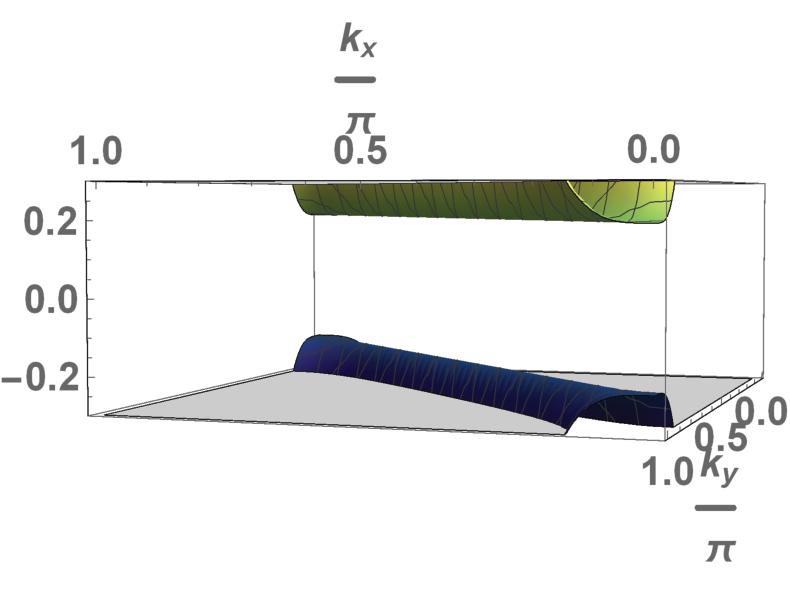} c)\\
                  }
\end{minipage}
\caption{(Color online)
The gapped fermion spectrum of the Hubbard chain  (a) (dotted lines mark the gap) and 2D Hubbard model (b) and
a low energy spectrum (c) as function of wave vector, calculated at $\lambda =0.2$.
The gap is qual to $2\lambda$.
  }
\label{fig:1}
\end{figure}

We will analyze the formation of a gap in the spectrum of fermions at half-filling and focus our attention on opening the gap with increasing interaction in the chain and  square, cubic lattices.
The abnormal average, that breaks  the conservation of the number of particles in each band and leads to breaking spontaneous symmetry, is described by the $\lambda$-field naturally introduced by the Hubbard-Stratonovich transformation (see subsection "methods"). Breaking symmetry leads to formation of a gap in the spectrum of fermions, to the metal-insulator Mott transition. The solution for $\lambda_j$ is determined by an unknown vector $\textbf{q}$, on which the energies of the quasiparticle excitations depend (8), where $\lambda_\textbf{j}=\exp(i \textbf{q} \textbf{j}) \lambda $. $q\neq 0$ lifts the degeneracy of the spectrum over the spin of electron  (splits into two bands), its value must be found from the minimum energy of system or action (7).  Due to symmetry of the spectrum (8), at half filling the chemical potential is equal to zero for arbitrary ${U}$ and $\textbf{q}$.
At T=0K the energy of the system is determined by the quasiparticle excitations as $E=\sum_{k,E_\pm(k)<0}[E_- (k)+E_+ (k)]$. Calculations show, that the minimum energy $E$ is achieved at $q=\pi$ for arbitrary $\lambda$ and dimension of the lattice ${d}$ (lattice constant equal to 1). When $\overrightarrow{q}=\overrightarrow{\pi}$, the gap in the spectrum of fermions opens (see in Figs \ref{fig:1}) (in other words $\overrightarrow{q}=\overrightarrow{\pi}$ triggers a gap), the energy decreases.
As result the energies of the quasiparticle excitations (8) are reduced to the following $E({k})=\pm \sqrt{\varepsilon^2_d ({k})+\lambda^2}$, where $\varepsilon_d ({k})=-2\sum_{i=1}^{d} \cos {k}_i$, ${k}_i=({k}_x,{k}_y,{k}_z)$, where $\varepsilon_d({k}+\pi)=-\varepsilon_{d}({k})$.  This solution for $q$ resembles $\eta$-pairing mechanism of electrons with momenta  $\overrightarrow{k}$ and $\overrightarrow{k} + \overrightarrow {\pi}$ (where $\overrightarrow{\pi}=(\pi,\pi,\pi)$),  proposed in \cite{Y} (see also \cite{Yu}),
which leads to the condensation of Cooper pairs with nonzero momentum. As noted by Yang \cite{Y}, $ \eta$-paring is characteristic of lattice models and is absent in the continuum. We can continue this idea - opening  a gap in the spectrum at a fixed filling is also a prerogative of lattice models, the considered approach cannot be realized in any continuum model. Such, low energy spectrum of the Yang-Gaudin Fermi gas  with contact interaction is gapless \cite{Y0,G} unlike the Hubbard chain \cite{LW}.

\begin{figure}[tp]
     \centering{\leavevmode}
\begin{minipage}[h]{.315\linewidth}
\center{
\includegraphics[width=\linewidth]{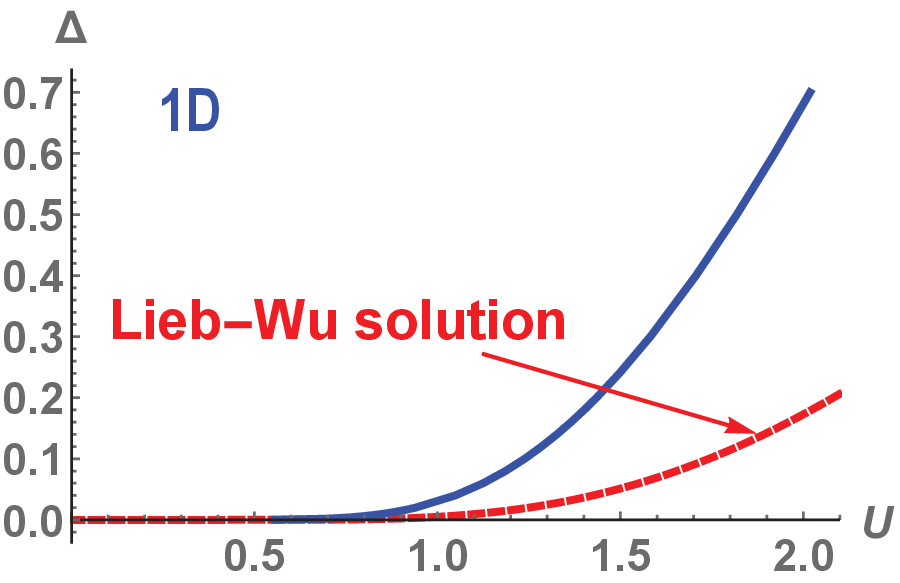} a)\\
                  }
    \end{minipage}
     \centering{\leavevmode}
\begin{minipage}[h]{.315\linewidth}
\center{
\includegraphics[width=\linewidth]{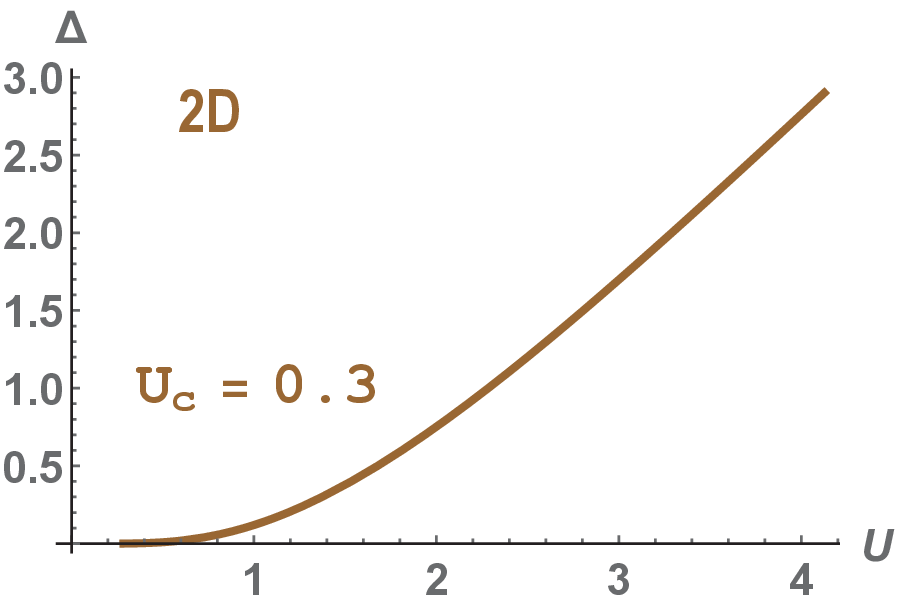} b)\\
}
 \end{minipage}
 \centering{\leavevmode}
\begin{minipage}[h]{.315\linewidth}
\center{
\includegraphics[width=\linewidth]{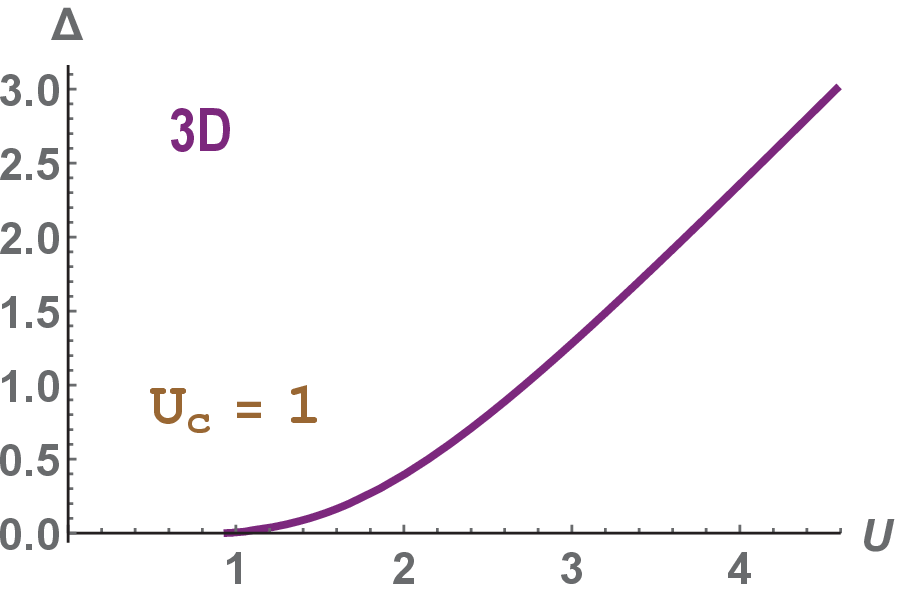} c)\\
                  }
    \end{minipage}
\caption{(Color online)
The gap $\Delta$ in the fermion spectrum as function of the strength of the on-site Hubbard interaction in the chain (a) (a red line shows the exact solution), square (b) and cubic (c) lattices.
  }
\label{fig:2}
\end{figure}

At half filling occupation an equation which corresponds to the minimal action $S_{eff}$
has the following form
\begin{eqnarray}
\frac{\lambda}{{U}}-\frac{{T}}{N}\sum_k \sum_{n} \frac{\lambda}{\omega^2_n+E^2({k})}=0,
\label{eq-2},
\end{eqnarray}
where N is the total number of atoms.

At T=0K nontrivial solution for $\lambda$ follows from equation $\frac{1}{{U}}=\frac{1}{N}\sum_k \frac{1}{2\sqrt{\varepsilon^2_d ({k})+\lambda^2}}$. In the ${U}\to \infty$ limit the gap is given by $\Delta ={U}$, since $\Delta=2\lambda$ (see in Figs 1, as an illustration), the result does not depend on the dimension of the lattice.

In the case of the Hubbard chain we can calculate the weak coupling limit ${U}\to 0$. We obtain the following expression for the gap $\Delta\simeq G \exp(-2\pi/{U})$, where the pre-exponential factor $G$ is determined by the integration region on ${k}$ near the Fermi energy ($\mu =0$).
According to exact solutions of the Hubbard chain
$\Delta = \frac{16}{{U}} \int_1^\infty d{x} \frac{\sqrt{{x}^2-1}}{\sinh(2\pi{x} /{U})}$ \cite{LW,AA}
and Heisenberg chain \cite{G}  the fermion spectra are gapped at half-filling for arbitrary on-site repulsion in the Hubbard chain and in the case of strong coupling in the Heisenberg chain \cite{G,AA1}. We can compare this result with the asymptotic expression at ${U}\to 0$ for the gap  $8\sqrt{{U}}/\pi \exp(-2\pi/{U})$ \cite{LW,AA} (see in Fig 2a).
\begin{figure}[tp]
     \centering{\leavevmode}
\begin{minipage}[h]{.47\linewidth}
\center{
\includegraphics[width=\linewidth]{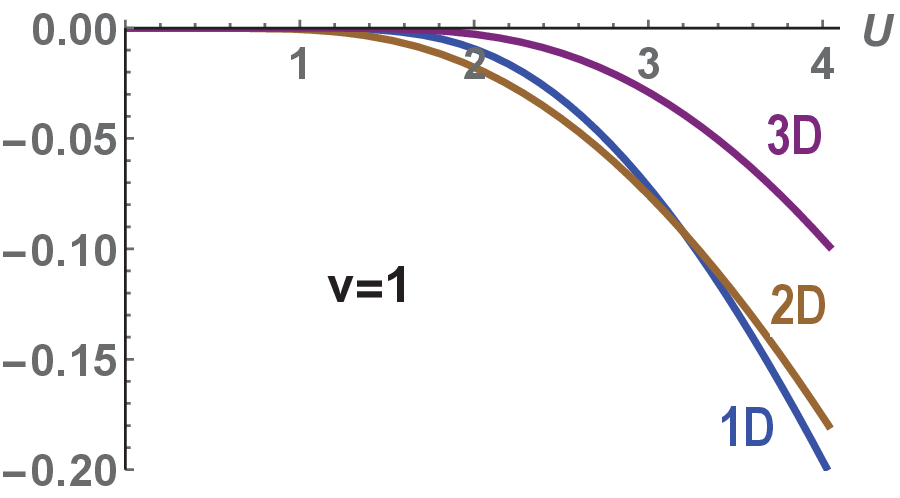} a)\\
                  }
    \end{minipage}
     \centering{\leavevmode}
\begin{minipage}[h]{.47\linewidth}
\center{
\includegraphics[width=\linewidth]{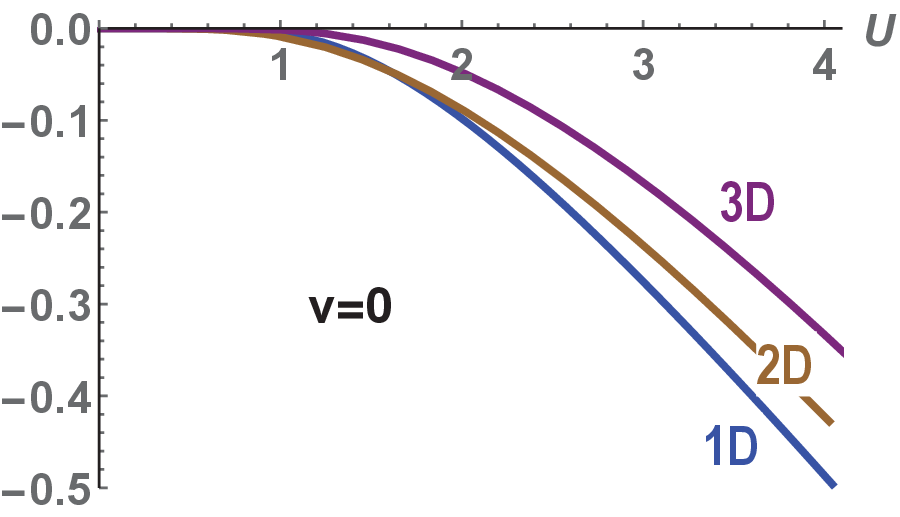} b)\\
}
 \end{minipage}
     \caption{(Color online)
$\delta S_{eff}$ as function of the strength of the on-site Hubbard interaction for the chain, square and cubic lattices calculated in two limiting cases for v=1 (the Hubbard model) and v=0 (the Falicov-Kimball model).
  }
\label{fig:3}
\end{figure}
In 2D and 3D model a quantum phase transition between metal and insulator states occurs at critical value of the on-site Hubbard interaction ${U}_c$,  when  the gap opens in the fermion spectrum. In Figs 2b, 2c we present results of numerical simulations for the gap in the spectrum of square and cubic lattices, respectively. Nontrivial solution for $\lambda$  (2) determines the value of the gap in the fermion spectrum.
Numerics performed supports numerical calculations, that in 2D and 3D Hubbard model the gap  in the spectrum opens at a finite value of the on-site Hubbard interaction, numerical calculation gives the following values ${U}_c=0.3$ for square and ${U}_c=1$ for cubic lattices. Should be noted that, as in all numerical calculations, the value of  $U_c$ increases with an increase in the dimension of the system. This is understandable at least because this increases the number of nearest neighbors or the band width. The calculated values of $U_c$ are less that numerical values of $U_c=2$ \cite {1} and $U_c=6$ \cite {3a}, as  noted above.

From numerical calculations it follows, that a nontrivial solution for $\lambda $ decreases the action, thus the phase with the gap in the spectrum is stable, it is the ground state.  Fig 3a  presents numerical calculations of the action $\delta S_{eff} = S_{eff} (\lambda)-S_{eff}(0)$ depending on the on-site Hubbard interaction for dimension d=1,2,3. The behavior of $\delta S_{eff}$ is same for different dimension, it is a monotonic decreasing function of ${U}$.

\section{Two band model of spinless fermions}
We consider same generalization of the Hubbard model, namely two band model of spinless fermions with different hopping integrals interacting via the on-site Coulomb repulsion. The  model Hamiltonian is written as
\begin{eqnarray}
&&
{\cal H}= - \sum_{<ij>} ({u} a^\dagger_{i}a_{j}+{ v }c^\dagger_{i}c_{j})-\mu \sum_{j}(n_{j}+m_j)+
%\nonumber\\&&
{U} \sum_{j}\left(n_{j}-\frac{1}{2}\right)\left(m_{j}-\frac{1}{2}\right),
\label{eq-H}
\end{eqnarray}
where $a^\dagger_{j}$ ($a_{j}$) and $c^\dagger_{j}$ ($c_{j}$) are the fermion operators that determine the spinless fermions of different bands on a site \emph{j},  $n_{j} =a^\dagger_{j}a_{j}$ and $m_{j} =c^\dagger_{j}c_{j}$ denote the density operators. The Hamiltonian (3) describes the hoppings of fermions between the nearest-neighbor lattice sites  with the magnitudes  $u$ and $v$, the on-site repulsion between fermions of different bands is taken into account via parameter ${U}>0$. At $u=v$ the model (3) reduces to the Hubbard model (1), where the orbital index is actually a spin index. In the case of one flat band (${v}=0$) the Hamiltonian (3) is reduced to the spinless Falicov-Kimball model, in which a flat band lies on the Fermi energy  \cite{DG1,DG2}. Note, that for non degenerate bands, when $u\neq v$, the (1+1)D model (3) is not integrable.

As in the previous subsection, we consider the case of a half-filled occupation in which $\mu=0$ is realized for arbitrary $\textbf{q}$, ${v}$ and dimension of the lattice. The energy calculations  performed for different ${v}$ and dimension of the model support result of the Hubbard model, that minimum of the energy is reached at $\textbf{q} = \overrightarrow{\pi}$.  The energies of quasiparticle excitations are determined as follows
\begin{eqnarray}
&&
E_{1\pm}({k})= \frac{1}{2} [ \varepsilon_{1d}({k}+\pi)+ \varepsilon_{2d}({k})
\pm
\sqrt{4\lambda^2+(\varepsilon_{1d}({k}+\pi)-\varepsilon_{2d}({k}))^2}],  \nonumber \\
&&
E_{2\pm}({k})= \frac{1}{2} [ \varepsilon_{2d}({k}+\pi)+\varepsilon_{1d}({k}) \pm
\sqrt{4\lambda^2+(\varepsilon_{2d}({k}+\pi)-\varepsilon_{1d}({k}))^2}
],
\label{eq-1}
\end{eqnarray}
where $\varepsilon_{1d}({k})= -2 u \sum_{j=1}^d \cos{k}_j$ and $\varepsilon_{2d}({k})= -2 v\sum_{j=1}^d \cos{k}_j$, it is convenient to put ${u}=1$ and consider $0\leq{v}<1$ (the case ${v}=1$ has been considered in the previous subsection).

At half filling occupation the spectrum is symmetric with respect to zero energy, it includes four branches (see in Figs 4). The gap in the fermion spectrum $\Delta$ is determined by the following expressions: $\Delta = 4 \lambda \frac{\sqrt{{v}}}{1 + {v}}$ for $\lambda\leq\lambda_0$  and  $\Delta =2 d({ v}-1) +2\sqrt{\lambda^2 + d^2 (1+{ v})^2}$ for $\lambda\geq\lambda_0$, where $\lambda_0 = 2 d \frac{\sqrt{{v}} (1 + {v})}{1-{v}}$. In the limiting cases, when ${v}=1$ and ${v}=0$, the gap is defined as follows $\Delta =2\lambda$ and $\Delta=-2d+2\sqrt{\lambda^2+d^2}$ for arbitrary $ \lambda$, since $\lambda_0 \to \infty$ at ${v} \to 1-0$ and $\lambda_0=0$ at ${v} =0$.

At T=0K an equation which corresponds to the minimal action $S_{eff}$ has the following form
\begin{eqnarray}
\frac{\lambda}{{U}}- \frac{1}{N}\sum_k \frac{\lambda}{\sqrt{4\lambda^2+(1+{v})^2 \varepsilon_d^2(k)}}
=0.
\label{eq-Eqs}
\end{eqnarray}

\begin{figure}[tp]
     \centering{\leavevmode}
\begin{minipage}[h]{.315\linewidth}
\center{
\includegraphics[width=\linewidth]{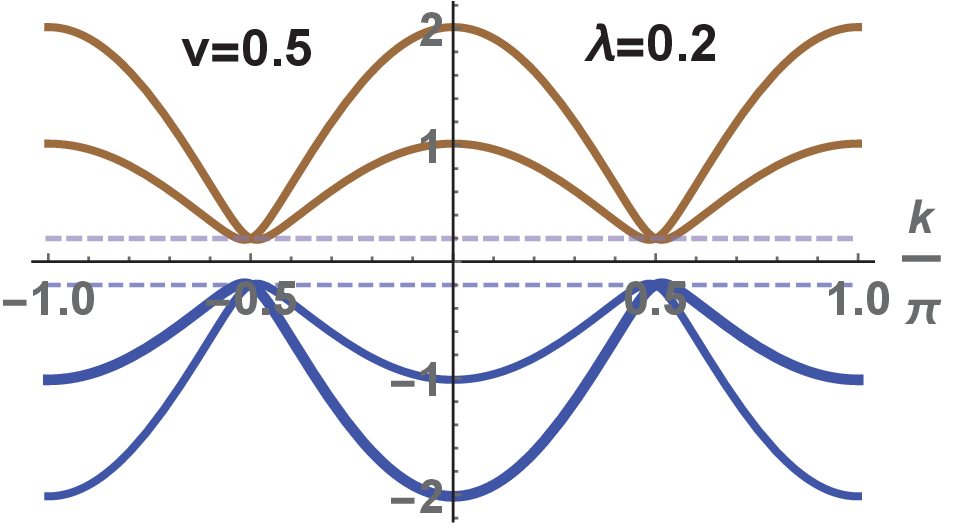} a)\\
                  }
    \end{minipage}
     \centering{\leavevmode}
\begin{minipage}[h]{.315\linewidth}
\center{
\includegraphics[width=\linewidth]{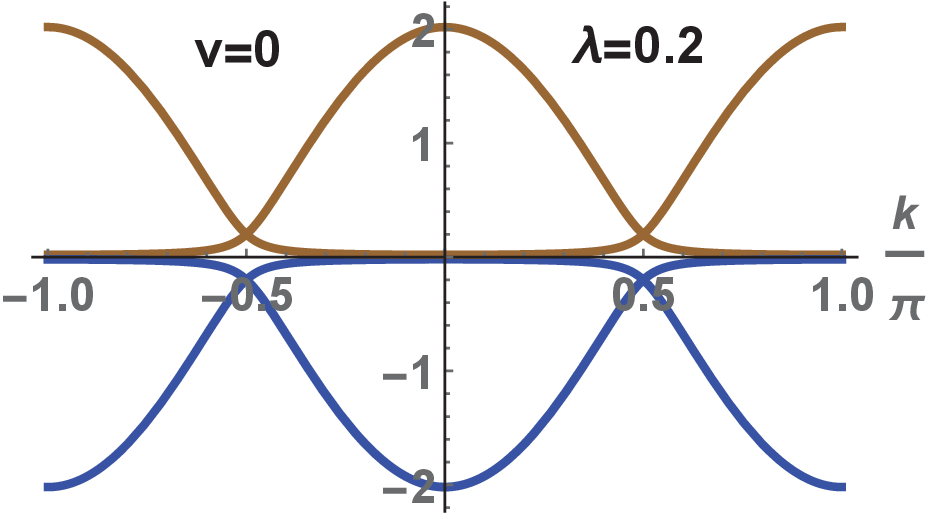} b)\\
}
 \end{minipage}
 \centering{\leavevmode}
\begin{minipage}[h]{.315\linewidth}
\center{
\includegraphics[width=\linewidth]{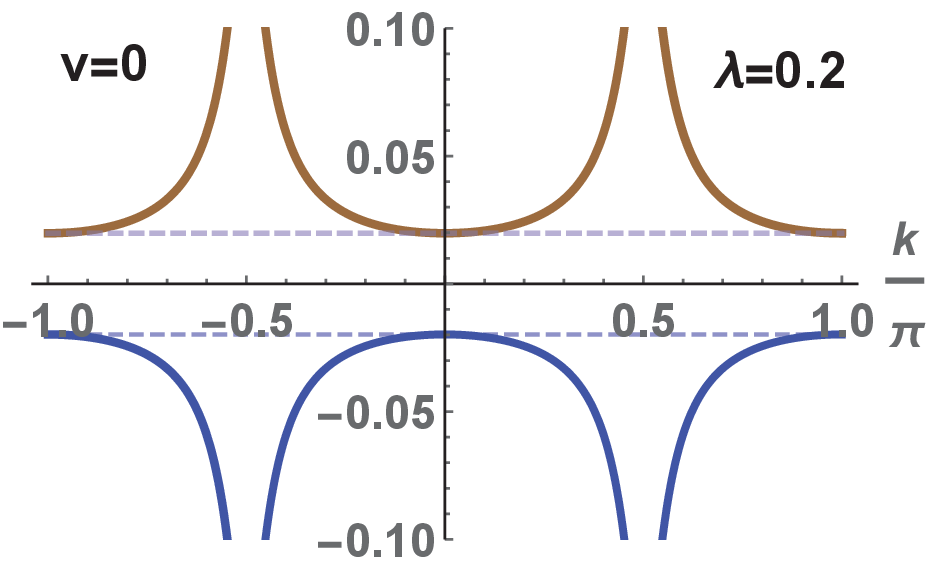} c)\\
                  }
    \end{minipage}
\caption{(Color online)
The spectrum of the chain of spinless fermions as function of wave vector calculated at $\lambda =0.2$ for ${v}=0.5$ (a) (dotted lines fix the gap, here $\Delta =\frac{4\lambda \sqrt{v}}{1+v}$) and ${v}=0$ (b), low energy spectrum (c) (here $\Delta=2\sqrt{1+\lambda^2}-2)$.
  }
\label{fig:4}
\end{figure}

To visualize the behavior of the gap in the model (3) we plot 3D graphics in Figs 5  for different dimensions of the system.
In the chain in the weak interaction limit, the gap in the spectrum is exponentially small $ \Delta \simeq G \exp {\left(-\frac{\pi\sqrt{1+v^2}}{U}\right)}$, like this takes place in the Hubbard model. The gap in the spectrum calculated in square and cubic lattices opens at finite values of on-site interaction, a minimal value of the gap is realized in the case of one flat band (when ${v}=0$). It should be noted that in the limit v=0, the vector $\textbf{q}$  is equal to $\pi$ for dispersion band and is not defined for states of the flat band, since its dispersion does not dependent on the wave vector. In the case $v<0$ $q=0$, the value of $q$ changes abruptly from $\pi$ to 0 at $v=0$.  This can lead to $\lambda$-field phase fluctuations, which are taken into account in \cite{DG2}.

\begin{figure}[tp]
     \centering{\leavevmode}
\begin{minipage}[h]{.4\linewidth}
\center{
\includegraphics[width=\linewidth]{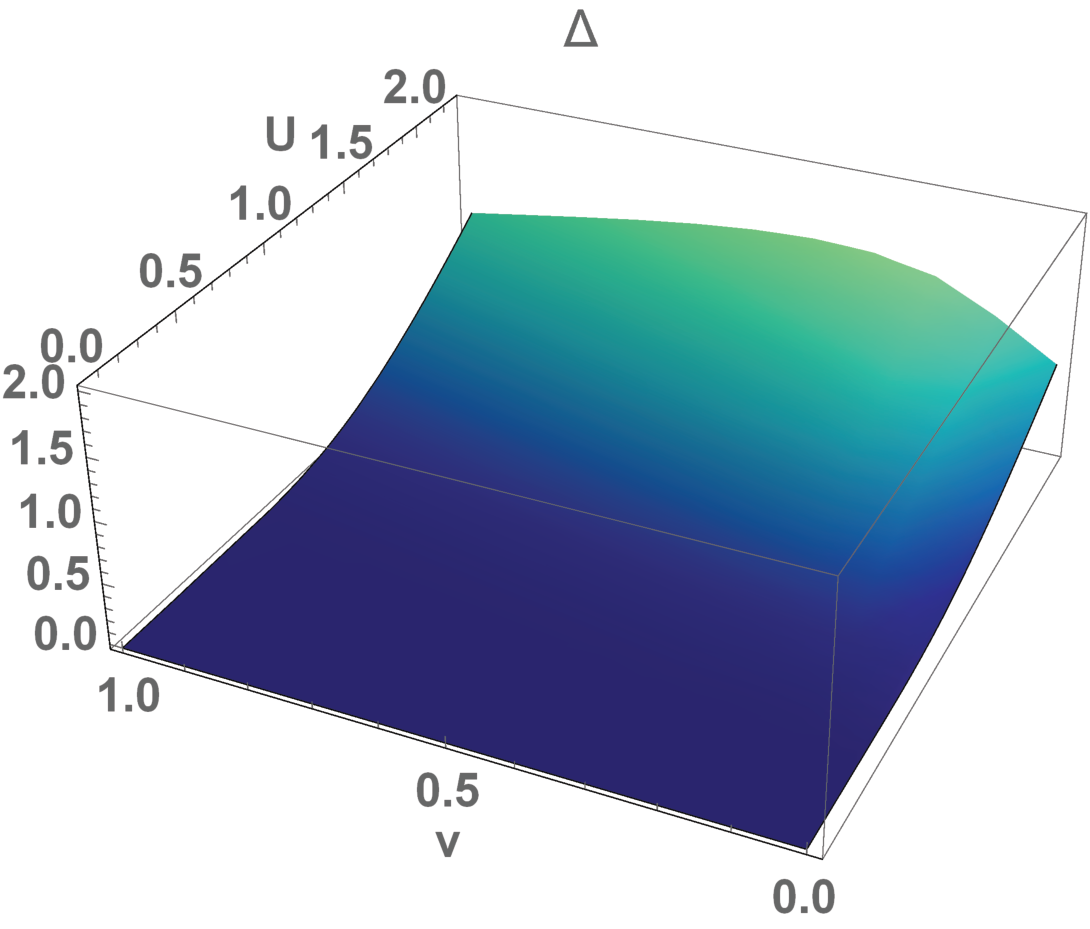} a)\\
                  }
    \end{minipage}
     \centering{\leavevmode}
\begin{minipage}[h]{.55\linewidth}
\center{
\includegraphics[width=\linewidth]{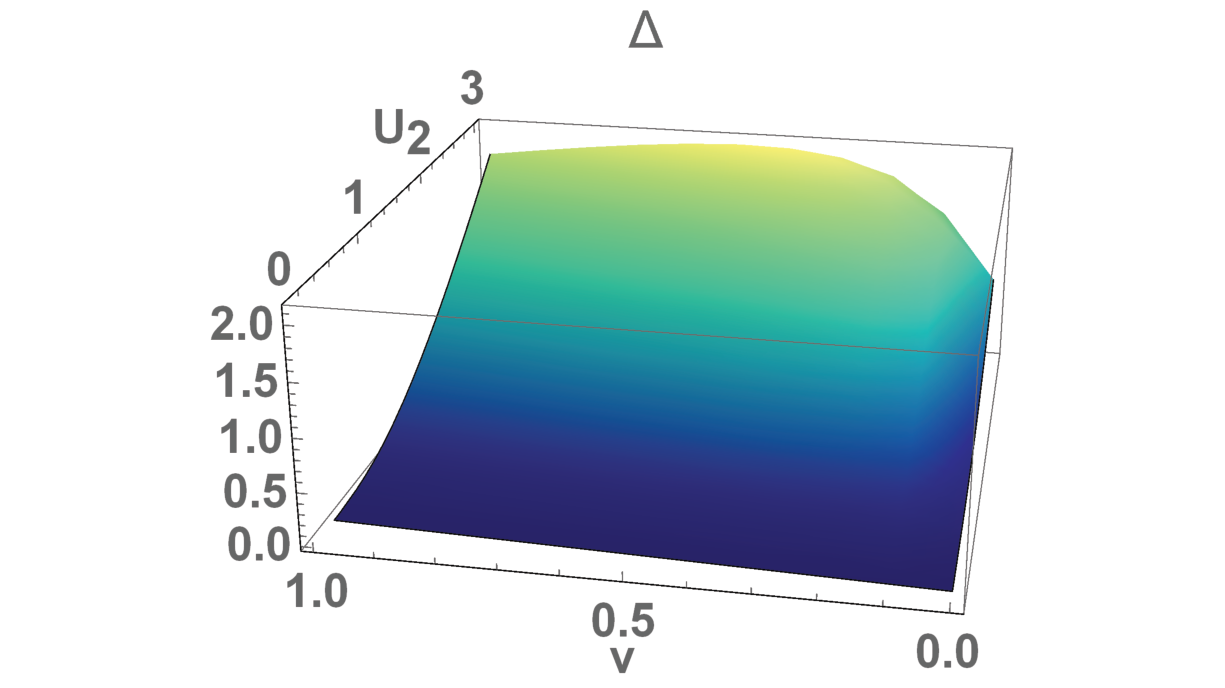} b)\\
}
 \end{minipage}
 \centering{\leavevmode}
\begin{minipage}[h]{.5\linewidth}
\center{
\includegraphics[width=\linewidth]{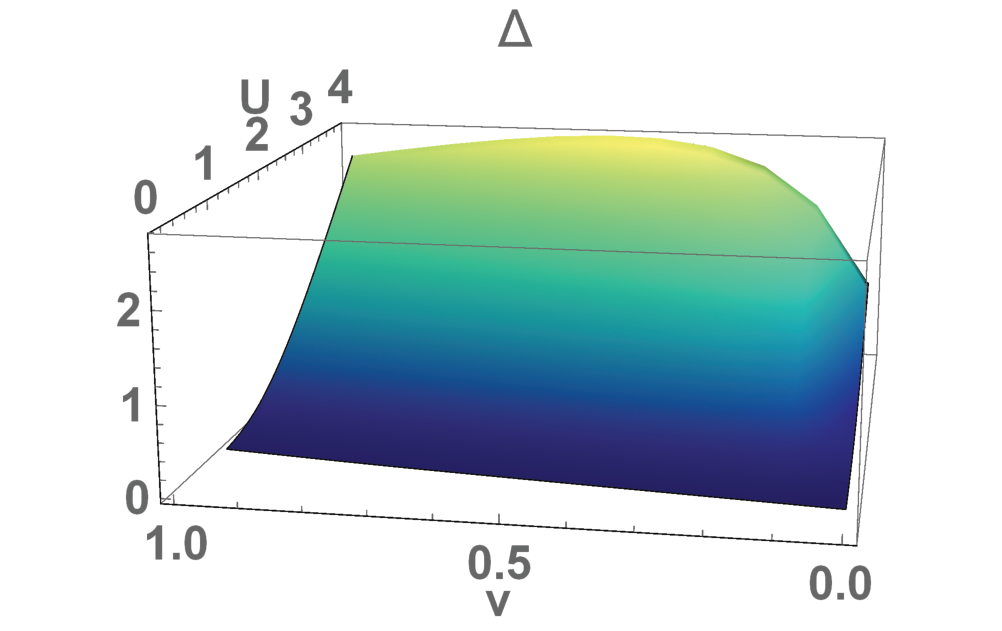} c)\\
                  }
    \end{minipage}
\caption{(Color online)
The gap $\Delta$ in the fermion spectrum as function of the strength of the on-site Hubbard interaction U and hopping integral v calculated at half filling for the chain (a), and square  (b), cubic  (c) lattices.
  }
\label{fig:5}
\end{figure}

\section{Conclusions}
We have shown that when half-filled occupation, fermions of different bands hybridize with momenta, shifted by $ \pi $, which  leads to the formation of a gap in the fermion spectrum. The gap state is stable, it corresponds to breaking spontaneous symmetry. It is shown that in the chain the gap occurs for arbitrary value of the on-site Coulomb interaction, in the weak interaction limit it is exponential small, as in an exact solution of the Hubbard chain. In 2D and 3D dimensions the insulator state is formed at a finite interaction value, so the Mott transition happens for 2D (square lattice) at ${U}_c=0.3$ and for 3D (cubic lattice) at ${U}_c=1$. The proposed approach allows us to describe the metal-insulator Mott transition in the Hubbard and Falicov-Kimball models in one formalism.

\section{Acknowledgments}
Author thanks Denis Golosov for interesting discussions.
The studies were also supported by the National Academy of Sciences of Ukraine within the budget program 6541239 "Support for the development of priority areas of scientific research".

\section{Methods}
\subsection{On-site Hubbard interaction}

The model Hamiltonian for spin-degenerate electronic states is defined as ${\cal H} = {\cal H}_{0}+{\cal H}_{int}$, where  ${\cal H}_{0} = - \sum_{<i,j>} \sum_{\sigma =\uparrow,\downarrow} a^\dagger_{i,\sigma} a_{j,\sigma}$ and ${\cal H}_{int}={U } \sum_{j}n_{j,\uparrow}n_{j,\downarrow}$.  The term ${\cal H}_{int}$ is conveniently rewritten as ${\cal H}_{int}= -{U}\sum_{j}\chi_j^\dagger\chi_j$, where $\chi_j^\dagger= a^\dagger_{j,\uparrow}a_{j,\downarrow}$. The Hubbard-Stratonovich transformation maps interacting fermion systems to non-interacting fermions moving in an effective field (the $\lambda$-field), we define the interaction term introducing the action ${S}_{0}$
\begin{equation}
{ S}_{int}={S}_{0}+ \sum_{j}\left(\frac{\lambda^\ast_{j}\lambda_{j}}{{U}}+\lambda_{j}\chi^\dagger_{j}+\lambda^\ast_{j}\chi_{j}\right).
 \label{A1}
\end{equation}

The canonical functional is defined as ${\cal Z}=\int {\cal D}[\lambda] \int {\cal D}[\chi\dagger,\chi] e^{-S}$, where the action $S=\frac{1}{U}\sum_{j}\lambda^\ast_{j}\lambda_{j}
+ \int_0^\beta d\tau \Psi^\dagger (\tau)[\partial_\tau  + {\cal H}_{eff}]\Psi (\tau)$ with
$ {\cal H}_{eff}=  {\cal H}_0 +\sum_{j}(\lambda_{j}\chi^\dagger_{j}+\lambda^\ast_{j}\chi_{j})$, where $\Psi (\tau)$ is the wave function. We expect $ \lambda_ {j} $ to be independent of $ \tau $ due to translational invariance.
According to the exact solution of the (1+1)D Hubbard model, the  on-site interaction does not break translational symmetry, therefore only the phase of the $ \lambda $-field in Eq.(6) depends on the wave vector, namely
$ \lambda_\textbf{j} = \exp (i \textbf {q} \textbf {j}) \lambda $. Below we consider this solution for $\lambda_\textbf{j}$, which takes place at local hybridization in (6). One can integrate out the fermionic contribution to obtain the action $S_{eff}$ per atom for the $\lambda$-field, where $\omega_n ={T}(2n+1)\pi$ are the Matsubara frequencies
\begin{equation}
\frac{S_{eff}}{\beta}=-\frac{{T}}{N}\sum_{\textbf{k}}\sum_n \ln [(-i\omega_n+E_+(\textbf{k}))(-i \omega_n+E_-(\textbf{k}))]+\frac{|\lambda|^2}{{U}},
 \label{A2}
\end{equation}
where
\begin{equation}
E_\pm({k})]=\mu + \frac{\varepsilon({k})+\varepsilon({k}+q)}{2} \pm\frac{1}{2}\sqrt{
4\lambda^2 + (\varepsilon({k})-\varepsilon({k}+q))^2}.
 \label{A3}
\end{equation}
It should be noted that the unit cell doubles, it is determined by two wave vectors  $\textbf{k}$ and $\textbf{k} + \textbf{q}$. We define a solution for $\lambda$ in the saddle point approximation for the functional $ {\cal L}$, the minimal action $ S_{eff} $ will dominate if $\lambda$ satisfies the following condition $\partial S_{eff}/\partial\lambda =0$.

%\section*{References}
%\bibliography{mybibfile}

\end{document}